\documentstyle[epsfig]{salva}
\begin{document}
\title{Review of $\alpha_s$ measurements at LEP 2}
\author{S. Mart\'{\i} i Garc\'{\i}a}
\address{Physics Department, University of Liverpool, Oxford St., 
Liverpool L69 7ZE, U.K. \\ e-mail: s.marti@cern.ch}
\maketitle
\begin{abstract}
  Since 1995,  LEP has steadily increased  the center of mass energy of
  the colliding beams,  from the $M_Z$  resonance to 133, 161 and 172
  GeV.  New measurements of the  strong coupling constant, $\alpha_s$,
  at these  energies have been performed  by  the LEP experiments, L3,
  ALEPH,  OPAL  and  DELPHI.  In  this  article,  the new  results are
  summarized, and combined  with  the   previous LEP measurement    of
  $\alpha_s (M_Z)$ in order to obtain an updated LEP average of
  \centerline{$\alpha_s (M_Z) = 0.120 \pm 0.005$.}
\end{abstract}
\section*{Introduction}
Over the past two years (1995 and 1996), the RF system of LEP has been
upgraded with the inclusion of new  superconducting RF cavities.  Thus
the energy of the LEP $e^-$ and $e^+$ beams has increased, giving
data with center of mass energy of 133\footnote{The true energies were
  130 and 136 GeV.  However, as the integrated luminosity at these two
  energy  points was the same,    the  data are usually combined   and
  presented as data collected at the mean energy of 133 GeV.}, 161 and
172  GeV.   The  analysis  of the   hadronic   system produced in  the
$e^+e^-\rightarrow Z/\gamma^* \rightarrow q\bar{q}$ process allows the
measurement of  the      running of the  strong   coupling   constant,
$\alpha_s$,  and the  test of QCD   predictions.  Any deviation can be
interpreted as   a  sign of  new  physics (e.g.   production  of light
gluinos) appearing at the new energy domain accessible at LEP 2.
\subsection*{Theoretical predictions}
It  is well known  that  $\alpha_s$  is  not a  fixed  but  a  running
quantity.  Its value depends  on the physical  energy scale ($\mu$) of
the process.  The beta  function  describes the renormalization  scale
dependence of the strong coupling constant:
\begin{equation}
\mu\frac{d\alpha_s(\mu)}{d\mu}= - \frac{\beta_0}{2\pi}\alpha_s^2(\mu) -
\frac{\beta_1}{4\pi^2}\alpha_s^3(\mu)+{\cal O}(\alpha_s^4)
\label{eq:beta}
\end{equation}

Assuming SU(3)  colour   gauge symmetry:   $\beta_0  = 11-2n_f/3$  and
$\beta_1=51-19n_f/3$,  with $n_f$   being the number   of active quark
flavours.  Solving equation  \ref{eq:beta},  the $\alpha_s$ dependence
with the scale goes as: $\ln(\mu^2/\Lambda^2)$, where $\Lambda$ is the
fundamental QCD scale.
\subsection*{Combination of individual results}
It is  a matter  of   discussion how  the different  measurements   of
$\alpha_s$  are  combined,   given that    most   of the   theoretical
uncertainties are  correlated \cite{schmelling}.  In this article, the
average  of a set of  measurements, $x_i\pm\sigma_i$, is calculated as
$\langle  x  \rangle = \Sigma  x_i  w_i$ (with weights $w_i$ inversely
proportional   to squares of the  error).    The experimental error is
obtained   in a similar way, while   the theoretical uncertainties are
taken in average.  The total error  corresponds with the quadratic sum
of the experimental and theoretical uncertainties.
\section*{Measurements of $\alpha_s$ at LEP 1}
The  high  statistics   collected  at  the Z  peak   by   the four LEP
experiments  (over  15 million  of hadronic   Z  decays)  has  allowed
$\alpha_s(M_Z)$  to be  measured  with  high  accuracy\footnote{Due to
  this,  most  of  the published   results  on $\alpha_s$   from other
  experiments at different energy  scales are extrapolated to a common
  reference scale, usually taken as $M_Z$.}.  All four LEP experiments
have published $\alpha_s$ measurements  with a wide  range of methods. 
These can be summarized by: inclusive quantities as $R_h$ and $R_\tau$
(the total hadronic decays of the $Z$ and the $\tau$), analysis of the
global  event    shapes  (thrust,   broadness...),  scaling violations
\cite{aleph_scaling,delphi_scaling},   three-jet rates, etc.   Another
set  of  important  measurements   was the  test   of the   $\alpha_s$
universality for all the  quark flavours \cite{delphi_ab,aleph_total}. 
Also  important are  the  measurements using $q\bar{q}\gamma$  events,
where  the boost to the  center of mass of  the hadronic system allows
measurements       of        the          running     of     $\alpha_s$
\cite{delphi_qg,marti,l3_qqgam}.

The global average of all LEP I measurements is: $\alpha_s(M_z)= 0.121
\pm  0.005$, where    the  main contribution  to   the  error  is  the
theoretical uncertainty.
\section*{Measurements of $\alpha_s$ at LEP 2}
The $\alpha_s$ measurements at LEP 2 have new problems with respect to
the measurements at LEP  1.  The first is  the limited statistics.  So
far, the  integrated luminosity of the  four  experiments in  LEP 2 is
$\approx 100$  pb$^{-1}$ (representing about  10,000 hadronic events). 
The second problem is the background, mainly from the radiative return
to the Z events,  and from $\gamma\gamma$  collisions. Another type of
background    comes    from  the    $W^+W^-$   pairs   decaying   into
$q\bar{q}q\bar{q}$ and $q\bar{q}l\nu$, however this background is only
present with LEP operating above the $W^+W^-$ production threshold. Due
to  these problems, the   total  sample available for each  individual
experiment and energy point is about 400$\sim$1000 events.

\subsection*{Global event shapes}
\begin{figure}[ht]
\centerline{\epsfig{file=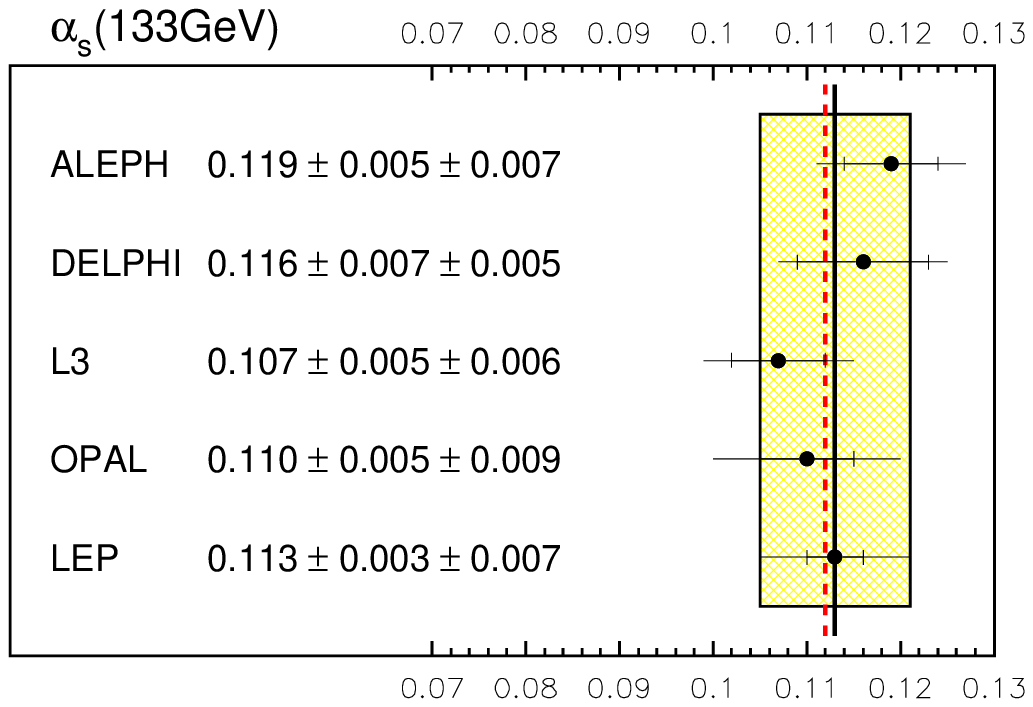,height=49mm}
\epsfig{file=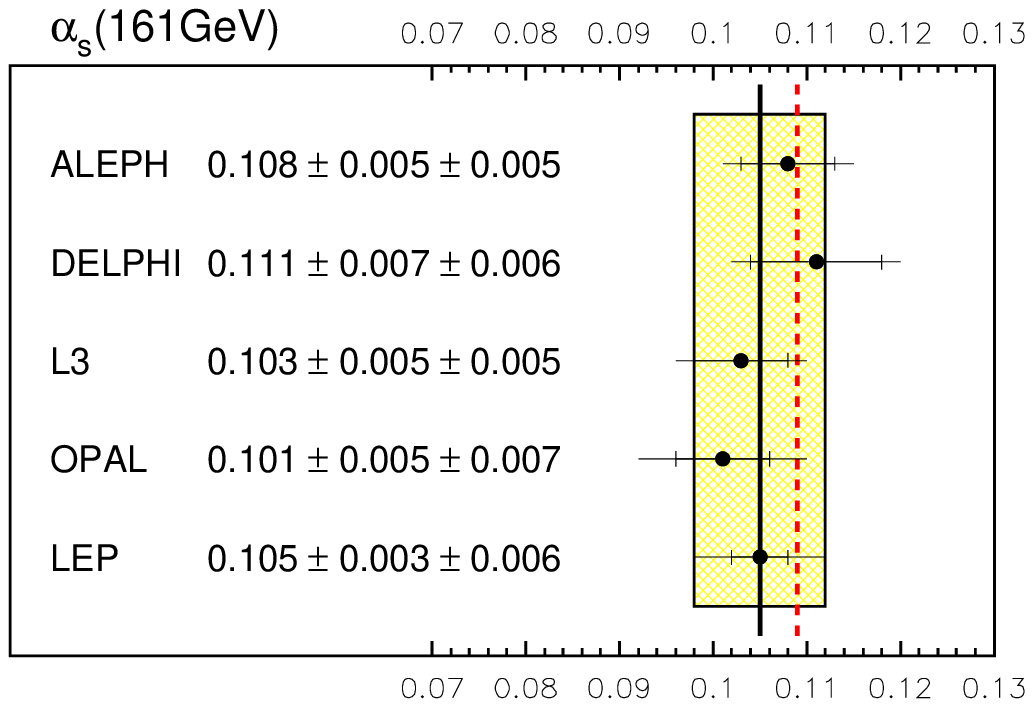,height=49mm}}
\centerline{\epsfig{file=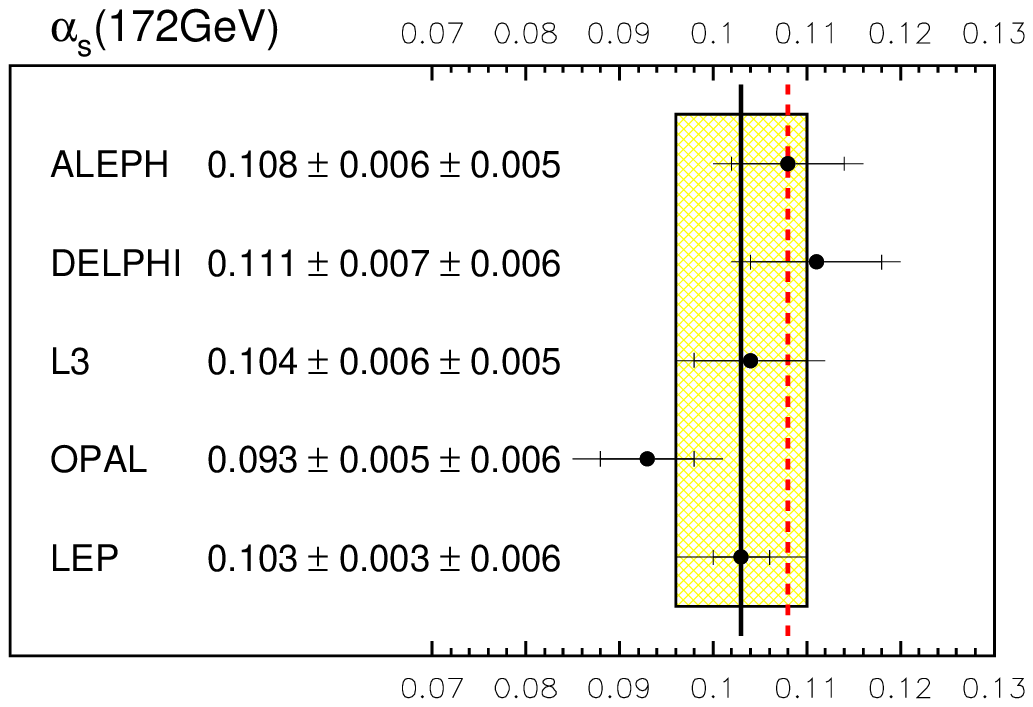,height=50mm}}
\caption{Measurements of $\alpha_s$ in LEP 2 at 133, 161 and 172 GeV. 
  For  each  measurement, the errors  quoted are  the experimental and
  theoretical respectively.    Dashed  lines  correspond  to   the QCD
  prediction assuming $\alpha_s(M_Z)=0.118$.}
\label{fig:lep2}
\end{figure}

The LEP experiments  obtained new  results on $\alpha_s(\sqrt{s})$  by
comparing   the global event   shape variables  with  the exact ${\cal
  O}(\alpha_s^2)$ QCD calculations plus either NLLA or resumed series.
The  new  LEP  2 results   for   $\alpha_s$ are  summarized  in figure
\ref{fig:lep2}.

The first   measurement  of $\alpha_s$(133GeV)  was   published  by L3
\cite{l3_133}.  In that  analysis the distributions of  thrust, scaled
heavy jet mass and jet broadening variables were compared with resumed
${\cal O}(\alpha_s^2)$  QCD calculations.  ALEPH \cite{aleph_133} used
the differential two  jet rate with the  Durham jet algorithm.  In the
DELPHI  analysis \cite{delphi_133}   a  similar  set  of event   shape
variables was used, plus  the three jet  rate with the Durham amd  the
JADE algorithms.      In  the  OPAL   analysis   \cite{opal_133}   the
distributions  of $1-T$,   scaled   heavy  jet  mass, jet   broadening
variables and the Durham differential two jet  rate were fitted to the
${\cal O}(\alpha_s^2)$+NLLA QCD predictions.  The same event variables
were used in the  analysis of  the 161 GeV  data, providing  the first
$\alpha_s$(161 GeV) LEP measurement \cite{opal_161}.

The results at 172 GeV shown in figure \ref{fig:lep2} were provided by
the QCD representatives of each experiment and are still preliminary.

The size of  the nonperturbative effects in  the event shape variables
decreases  when increasing  the  energy, as   shown in  a DELPHI study
\cite{delphi_133,klaus}.  This is due to the convergence of the ${\cal
  O}(\alpha_s^2)$ perturbative calculations with the   nonperturbative
contribution described by a power series of $\alpha_s$.

\subsection*{LEP average}
\begin{figure}[hb]
\centerline{\epsfig{file=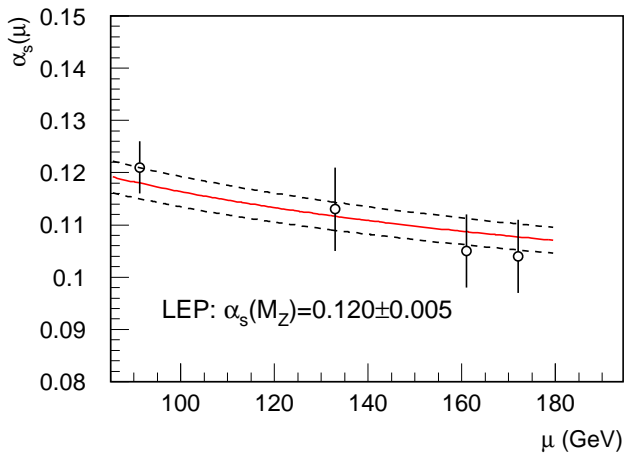,height=65mm}}
\caption{Running of the strong coupling constant: measurements at different 
  scales in LEP ($M_Z$,  133, 161 and 172  GeV  by order).  Only total
  errors are displayed.}
\label{fig:all_lep}
\end{figure}

In figure \ref{fig:all_lep}, the  global LEP values for $\alpha_s$ are
presented for each energy available at  LEP.  The lines corresponds to
the running of $\alpha_s$ as obtained assuming $\alpha_s(M_Z)=0.118\pm
0.003$  and   using the  renormalization    group  equation.  The  new
measurements are compatible  with  the expected values of  $\alpha_s$. 
Combining all values  of $\alpha_s(M_Z)$ measured so far  at LEP 1 and
LEP 2, the new LEP average is:
$$\alpha_s(M_Z) = 0.120 \pm 0.005 $$

\section*{Comparison with other experiments}
\begin{figure}[ht]
\centerline{\epsfig{file=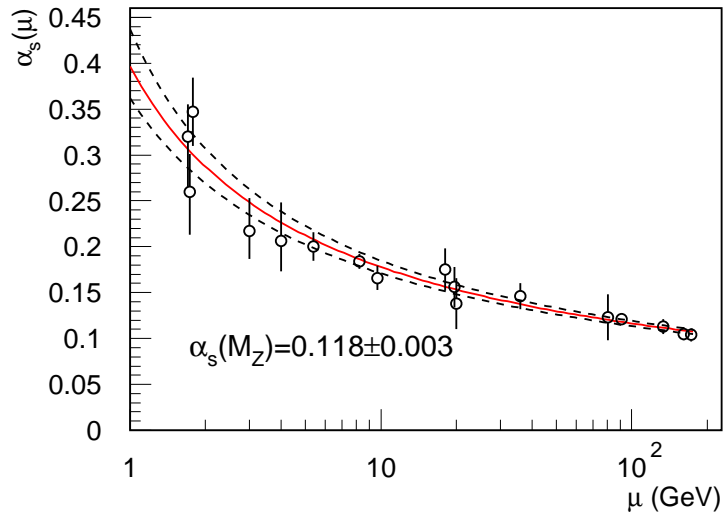,height=75mm}}
\caption{Running of the strong coupling constant: measurements at different 
  scales  compared to the  QCD prediction for $\alpha_s(M_Z)=0.118 \pm
  0.003$.   The  last three   points  correspond  to   the  new  LEP 2
  measurements. Only total errors are displayed.}
\label{fig:all_as}
\end{figure}

The LEP averages for  $\alpha_s$ at 133, 161 and  172 GeV can be added
to the plot showing  $\alpha_s$ running (figure \ref{fig:all_as}) over
a  wide range  of scales \cite{schmelling}.   With  the new $\alpha_s$
results from LEP 2, the global average is found to be the same than in
reference \cite{schmelling}:
\[ \alpha_s(M_Z) = 0.118 \pm 0.003\]

\subsection*{Acknowledgements}
I would like to thank G. Cowan, D. Duchesneau, J.  Fuster, L. del Pozo
and  D. Wicke  for their help  in preparing  the  talk.  Also to J.M.  
Hern\'andez, M.  Mart\'{\i}nez, J. Terr\'on  and  J.  Rold\'an for the
pleasant week in Chicago. I  apologise to everybody whose contribution
in this vast field I failed to mention.

\end{document}